\documentclass{iau-JDSS}
\usepackage{graphicx}

\title[Stellar and wind parameters] %% give here short title %%
{SpS5 - II. Stellar and wind parameters}

\author[short author list]   %% give here short author list %%
{F. Martins$^1$,
 M. Bergemann$^2$,
 J.M. Bestenlehner$^3$,
 P.A. Crowther$^4$,
 W.R. Hamann$^5$,
 F. Najarro$^6$,
 M.F. Nieva$^7$,
 N. Przybilla$^7$,
 J. Freimanis$^8$,
 W. Hou$^9$,
 L. Kaper$^{10}$,
 C.-D. Lee$^{11}$
}

\affiliation{$^1$LUPM, CNRS \& Universit\'e Montpellier II, Place Eug\`ene Batailon, F-34095, Montpellier, email: fabrice.martins@univ-montp2.fr\\[\affilskip]
$^2$Max-Planck-Institute for Astrophysics, Karl-Schwarzschild-Str.1, D-85741 Garching, Germany\\[\affilskip]
$^3$Armagh Observatory, College Hill, Armagh BT61 9DG, United Kingdom\\[\affilskip]
$^4$Department of Physics and Astronomy, University of Sheffield, Hounsfield Road, Sheffield, United Kingdom, S3 7RH\\[\affilskip]
$^5$Lehrstuhl Astrophysik der Universitaet Potsdam, Am Neuen Palais 10, 14469 Potsdam, Germany\\[\affilskip]
$^6$Centro de Astrobiolog\'ia, (CSIC-INTA), Ctra. Torrej\'on Ajalvir km 4, 28850 Torrej\'on de Ardoz, Spain\\[\affilskip]
$^7$Dr. Karl Remeis-Observatory Bamberg \& ECAP, University Erlangen-Nuremberg, Sternwartstr. 7, D-96049 Bamberg, Germany\\[\affilskip]
$^8$Ventspils International Radio Astronomy Centre, Ventspils University College, Inzenieru iela 101a, LV-3600 Ventspils, Latvia\\[\affilskip]
$^9$The National Astronomical observatories, Chinese Academy of Science\\[\affilskip]
$^{10}$Astronomical Institute Anton Pannekoek, University of Amsterdam, Science Park 904, 1098 XH Amsterdam, The Netherlands\\[\affilskip]
$^{11}$Graduate Institute of Astronomy, National Central University, Taiwan
}

\pubyear{2012}
\volume{Volume 16}  %% insert here IAU Highlights of Astronomy Volume Number
\pagerange{2--9}
\date{?? and in revised form ??}
\jname{Highlights of Astronomy, Volume 16}
\editors{T. Montmerle, ed.}
\begin{document}

\maketitle

\begin{abstract}
The development of infrared observational facilities has revealed a number of massive stars in obscured environments throughout the Milky Way and beyond. The determination of their stellar and wind properties from infrared diagnostics is thus required to take full advantage of the wealth of observations available in the near and mid infrared. However, the task is challenging. This session addressed some of the problems encountered and showed the limitations and successes of infrared studies of massive stars.
\end{abstract}

\firstsection % if your document starts with a section,

%-----------------------------------------------------------------------------------
\section{Introduction}

The use of IR for deriving stellar and wind parameters has become a necessity, as many stars are obscured. Many diagnostics are available in this range, however, so that using IR does not necessarily imply restrictions of the output quality. Line EWs and the appearance of the spectra can be used to derive temperature or spectral types, though morphological spectral types so inferred are less precise. This range is also helpful for mass-loss diagnostics, especially for low-mass loss rates. Two caveats, however, must be noted: the NIR lines form in the wind acceleration zone and they are very sensitive to NLTE  and 3D effects - they are thus extremely sensitive to modelling details. For most of the cases, observing K-band is sufficient, but diagnostics improve with J and H spectra, or - even better - if IR is complemented by optical and UV data (terminal velocities, are not constrained by IR, for example). Forbidden lines observed in the IR may arise from outer wind regions, so that also constitute sensitive problems of these poorly known zones. 

In this context, an improved knowledge of the atomic and molecule parameters is needed. In addition, 3D wind modelling may become necessary, to take into account both small-scale structures (clumps) and large-scale features (magnetically-compressed winds, wind collisions, corotating regions...). Convection in red supergiants also requires the use of 3D atmosphere models.

%-----------------------------------------------------------------------------------
\section{Stellar analysis from the infrared wavelength range}

Infrared analysis of massive stars is mandatory as long as there is a substantial amount of foreground extinction. In that case, the optical and UV flux are much lower than the infrared ones, despite the fact that hot massive stars emit most of their radiation at short wavelength. But infrared photometry and spectroscopy are also useful even if optical/UV data can be obtained, especially for luminosity and extinction determination. This is usually done by fitting the spectral energy distribution of synthetic models to flux-calibrated spectra and/or photometry (e.g., \cite{barniske08,paul06}). When infrared spectrophotometry is available, a much better accuracy in the luminosity determination can be obtained. The drawback is that uncertainties in photometry lead to larger errors on the luminosity in the infrared than in the optical (Bestenlehner et al., 2011 and this session). For red supergiants, infrared spectroscopy is extremely useful since these stars have their emission peak at those wavelengths.

Obtaining quantitative information on the stellar and wind parameters of obscured objects requires the use of quantitative analysis of infrared spectra. Historically, it is the near--infrared range (especially the H and K bands) which is favoured, since the contribution from thermal emission of the surrounding medium is weak at those wavelengths and increases at longer wavelengths. N. Przybilla presented a detailed view of the difficulties of modelling infrared spectra.

\begin{figure}[]
\centering
\includegraphics[width=13cm]{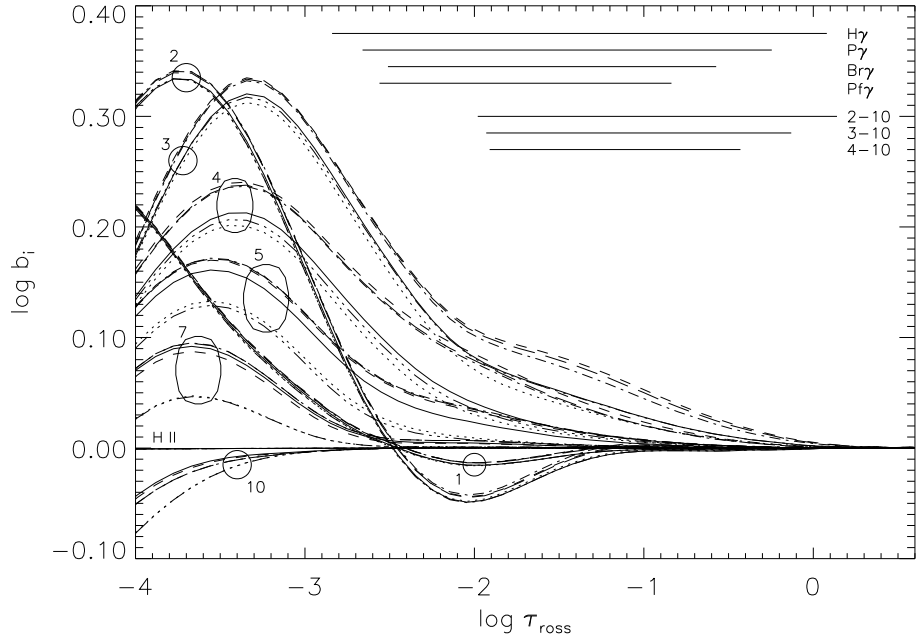}
%\caption{}
\caption{Effect of model atoms on departure coefficients of levels 1--5, 7 and 10 of hydrogen. The levels are indicated by their principal quantum number and are marked by circles. The line formation depth of several transition are indicated by the horizontal solid lines. From Przybilla $\&$ Butler (2004). } \label{pb04_bi}
\end{figure}

The first problem is related to non-LTE effects. Given the form of the source function, its variation with changes in the departure coefficients can be expressed as follows:

\begin{equation}
| \Delta (S_{l}) | = | \frac{S_{l}}{b_{i}/b_{j}-e^{-\frac{h\nu}{kT}}} \Delta (b_{i}/b_{j}) |
\end{equation}

\begin{figure}[]
\centering
\includegraphics[width=13cm]{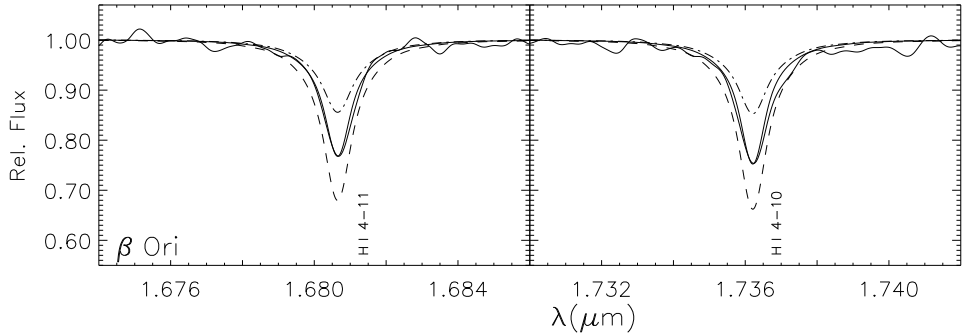}
%\caption{}
\caption{Brackett line profiles computed under LTE (dot-dashed lines) and non-LTE conditions with different collisional excitation prescriptions (dashed: approximate, according to Johnson 1972; dotted: data from new {\em ab-initio} computations). From \cite{pb04}. } \label{pb04_line}
\end{figure}

\noindent We see that at long wavelengths, the ratio $h\nu/kT$ becomes low. In that case, it can be shown that the source function is much more sensitive to changes in the departure coefficients than at optical or UV wavelengths. This implies that model atmosphere must account for non-LTE effects in great detail to correctly reproduce line profiles. Fig.\ \ref{pb04_bi} illustrates how different model atoms impact on the departure coefficients of a model with fixed stellar parameters. A simple change of the detailed input atomic data in the models has a
significant effect on the departure coefficients, and thus on the line
profiles. In Fig.\ \ref{pb04_line}, the profiles of Brackett lines are shown. First, the non-LTE effects are clearly seen: the LTE model produces much less absorption that the non-LTE models. But there is also a significant degree of variation among the non-LTE models depending on the type of collisional excitation formalism used. This stresses the need for accurate atomic data, even for hydrogen, to correctly model infrared spectra. For red supergiants, M. Bergemann showed that non-LTE effects are at least as important as for hot massive stars. A correct treatment of the radiative transfer leads to significant variations of the line profiles\footnote{Related to this topic, J. Freimanis presented a poster dedicated to polarized continuum radiative transfer in various nontrivial astrophysical coordinate systems, especially relevant for circumstellar gas envelopes.}.

\begin{figure}[]
\centering
\includegraphics[width=10cm,angle=-90]{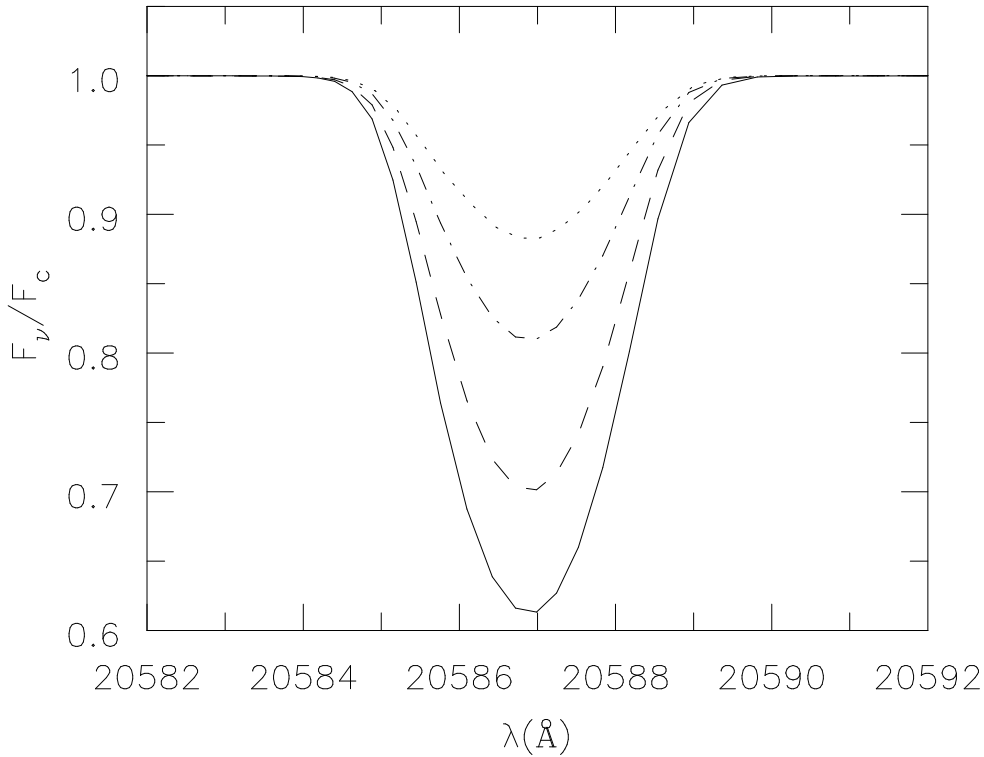}
%\caption{}
\caption{Effect of FeIV 584 \AA\ on HeI 2.058 $\mu$m. The solid line is the initial model. The dashed (dot-dashed, dotted) line corresponds to the same model in which the oscillator strength of FeIV 584 \AA\ has been reduced by a factor 2 (5, 10). From Najarro et al.\ (2006). } \label{paco06_heiIR}
\end{figure}

Given the sensitivity of infrared lines to non-LTE effects and model atoms, it is not surprising to see that line--blanketing effects are also extremely important. N. Przybilla illustrated the effect of line-blocking on the shape of the HeI 1.083 $\mu$m line. The emission is reduced by almost a factor of two when line-blocking is included, due to the different level populations and consequently different line source function. \cite{paco06} showed that even the very detail of atomic physics of minor FeIV lines in the extreme UV could affect the shape of the HeI triplet lines, in particular HeI 2.058 $\mu$m. This is illustrated in Fig.\ \ref{paco06_heiIR}. When reducing the oscillator strength of FeIV 584 \AA, the absorption of HeI 2.058 $\mu$m decreases. This is caused by the higher and higher population of the upper level of HeI 2.058 $\mu$m from which photons emitted in the HeI 584 \AA\ line are not stolen by the FeIV 584 \AA\ transition.

%-----------------------------------------------------------------------------------
\section{Key diagnostics}

The use of infrared spectrophotometry is helpful to constrain the spectral energy distribution, and thus the luminosity and extinction. But it is the analysis of medium--high resolution spectroscopy that provides information on the other stellar and wind parameters.

\smallskip

\underline{Effective temperature}: as at other wavelength ranges, the ionization balance is the most widely used method to constrain Teff. The HeI/HeII ratio is used when possible. W.R. Hamann pointed out in this session that the K-band is usually sufficient due to the presence of several HeI lines as well as HeII 2.189 $\mu$m. But below about 30000 K, the latter line disappears, and HeII lines present in the H and J bands need to be used. They also disappear rapidly when Teff decreases. N. Przybilla showed that for early B stars, the CII/CIII ionization ratio could be used, taking advantage of CII 0.9903 $\mu$m and CIII 1.1981--1.1987 $\mu$m (see also Fig.\ \ref{nieva09_tausco}). Several examples of temperature determinations have been shown by J. Bestenlehner, N. Przybilla, W.R. Hamann, P. Crowther. N. Przybilla highlighted that comparison of parameters derived purely from infrared diagnostics are usually in agreement with those resulting from optical studies, a conclusion also reached by \cite{repolust05}. N. Przybilla also indicated that for B stars, hydrogen lines can serve as secondary temperature indicators. M. Bergemann discussed different methods to determine T$_{\rm eff}$ for red supergiants, demonstrating that TiO bands severely underestimate temperatures when modeled in 1D LTE compared to more accurate full SED fits (Davies et al, in prep.).

\smallskip

\underline{Gravity}: hydrogen lines of the Brackett and Pfund series are the preferred diagnostics. In general not only their wings (as Balmer lines) but also their line core vary with $\log$ g. 

\smallskip

\underline{Surface abundances}: H and He lines present in the JHK bands provide constraints on the He/H ratio. This is true for OB stars, but more importantly for Wolf-Rayet stars which shows significant He enrichment. W.R. Hamann and N. Przybilla illustrated how synthetic spectra could be used to derive the helium content. N. Przybilla also highlighted that for A supergiants, abundances for C, N, O, Mg, Si and Fe could be derived from high-resolution spectroscopy. For O stars, nitrogen and in some cases carbon and oxygen abundances can be obtained. In Wolf-Rayet stars, several CIII, CIV, NIII lines give access to carbon and nitrogen abundances (as shown by F. Martins in session 1). P. Crowther presented determination of oxygen and neon abundances in WC stars using mid-infrared data (\cite{dessart00}, Crowther et al., in prep). The Ne content seems consistent with prediction of stellar evolution, while there might be a problem with oxygen. Abundances of Mg, Ti, Si, and Fe can be obtained from the J-band spectra of red supergiants. The lines of neutral atoms of these elements are very strong, and can be reliably studied even in low-resolution spectra. However, the non-LTE effects in these lines, especially Si I and Ti I, are very large (from -0.4 to +0.3 dex), requiring that non-LTE is properly taken into account (\cite{bergemann12}). The relatively strong emission lines of Mg, Si, Na and Fe in cool LBV stars are also used to derive metallicities (N. Przybilla, F. Najarro).

\smallskip

\begin{figure}[]
\centering
\includegraphics[width=13cm]{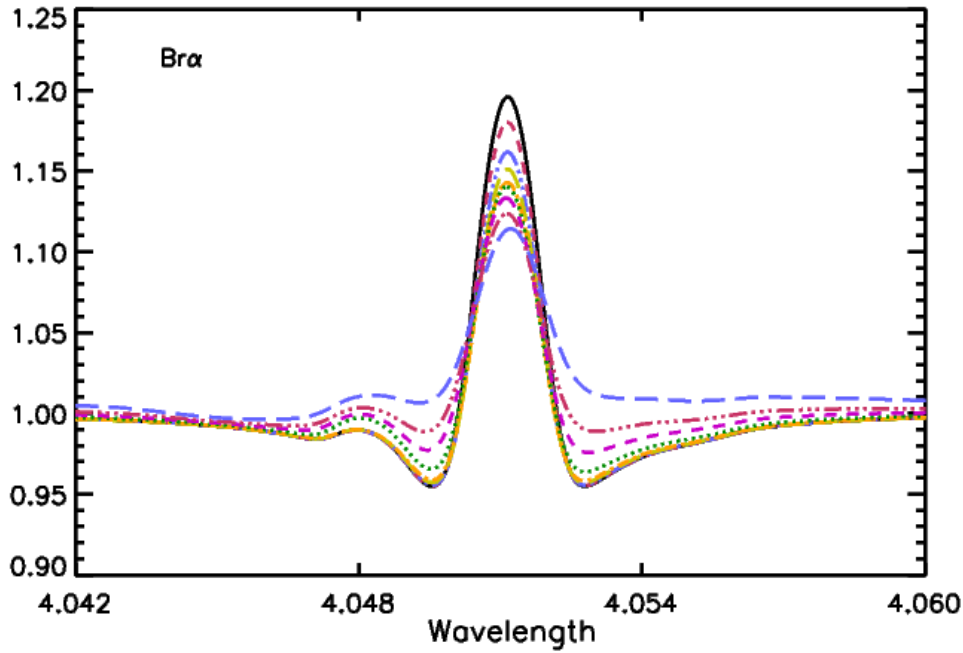}
\caption{Sensitivity of Br$\alpha$ to mass loss rate. The dashed line is a model with $\dot{M} = 1.0 \times 10^{-7}$ M$_{\odot}$ yr$^{-1}$ while the solid line is a model with  $\dot{M} = 5.0 \times 10^{-10}$ M$_{\odot}$ yr$^{-1}$. Other lines corresponds to models with intermediate mass loss rates. From \cite{paco11}. } \label{paco11_bra}
\end{figure}

\underline{Mass loss rate}: H and He emission lines of Wolf--Rayet stars are good diagnostics of the wind density and thus of mass loss rate (J. Bestenlehner, W.R. Hamann, P. Crowther). For OB stars, F. Najarro introduced Br$\alpha$ as the best mass loss rate indicator (\cite{paco11}). Other Brackett lines have been used (especially Br$\gamma$) but Br$\alpha$ reacts to density changes even for very low mass loss rates (in the so-called ``weak wind'' regime). The line core getting \textit{stronger} in emission when mass loss rate decreases, while the line wings are weakening. This peculiar behaviour is illustrated in Fig.\ \ref{paco11_bra}.

\smallskip

\underline{Clumping}: F. Najarro showed that the Bracket and Pfund hydrogen lines could be used to constrain the clumping distribution. In particular, the combination of Pf$\gamma$ formed in the inner wind and Br$\gamma$ emitted at intermediate depth is a powerful tool to constrain the clumping law in dense wind O stars. P. Crowther presented similar conclusions for Wolf-Rayet stars.

\smallskip

Several diagnostics of the stellar and winds properties of all types of massive stars are thus available. They are used for pioneering and systematic studies of massive stars in obscured environments.

%-----------------------------------------------------------------------------------
\section{Analysis of various types of massive stars}

During session 2, several examples of analysis of massive stars from infrared data have been presented. They are listed below.

\underline{A Supergiants}: N. Przybilla showed pioneering studies of A supergiants using high resolution spectroscopy obtained with CRIRES on the ESO/VLT. The results are presented in \cite{przy09} and Przybilla et al.\ (in prep). He first introduced the results of an analysis based on optical spectra to get a set of stellar parameters for Galactic A supergiants\footnote{Hou et al.\ presented a poster on the optical analysis of A stars.}. He then presented the results of the near-infrared analysis based on CRIRES spectra, highlighting that similar sets of parameters are found. This pilot study shows that future analysis of A supergiants in external galaxies, accessible with the new generation extremely large telescopes, will be feasible. This will provide information not only on the stars themselves, but also on galactochemical evolution and distance scales.
\smallskip

\begin{figure}[]
\centering
\includegraphics[width=13cm]{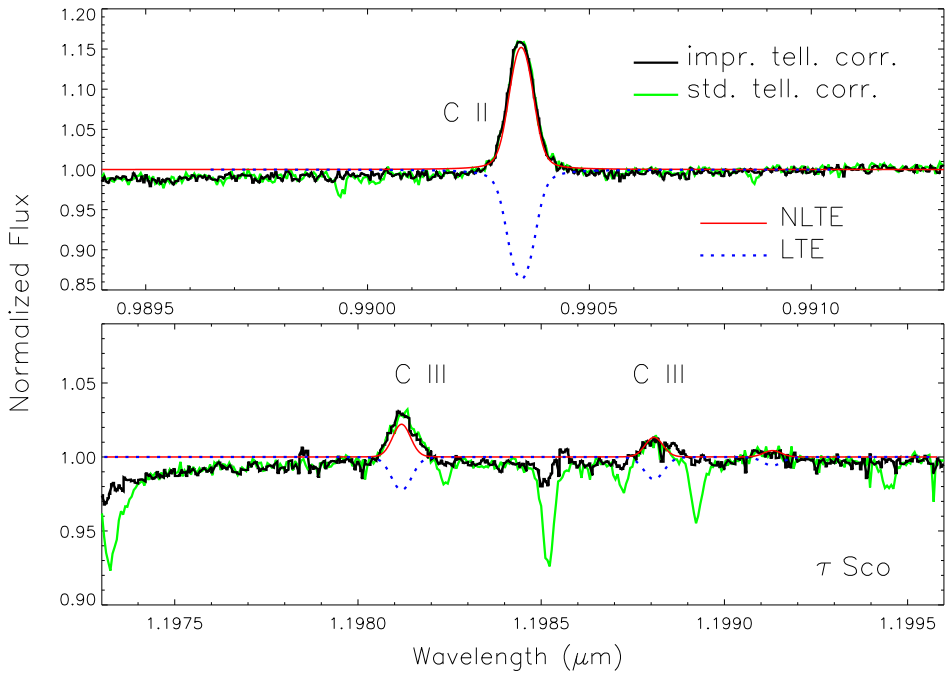}
\caption{Near-infrared carbon lines of the early B star $\tau$~Sco (black and green) together with a LTE synthetic spectrum (blue dots) and a non-LTE synthetic model (red solid). The use of non-LTE models is mandatory. The ratio of CII to CIII can be used to constrain the effective temperature. From \cite{nieva11}. } \label{nieva09_tausco}
\end{figure}

\underline{OB stars}: M.F. Nieva also presented analysis of CRIRES spectra of a few B stars, again pointing that results consistent with the optical are found (\cite{nieva11}). Fig.~\ref{nieva09_tausco} shows an example of the fit of carbon lines in the early B star $\tau$~Sco. Good agreement between model and observation is found if the LTE assumption is dropped. F. Najarro used Brackett and Pfund lines in several weak and dense wind O stars to constrain the clumping distribution and the mass loss rates. He confirmed previous results that late O dwarfs have mass loss rates as low as a few $10^{-10}$ M$_{\odot}$ yr$^{-1}$. For the dense wind star Cyg OB2 \# 7, he concluded that the clumping law was probably constant throughout most of the wind. Lee et al.\ presented preliminary results of the analysis of the variability of a Be star from infrared photometry and spectroscopy. Ellerbroek \& Kaper presented VLT/X-shooter spectroscopy of young massive stars.

\smallskip

\begin{figure}[]
\centering
\includegraphics[width=12cm]{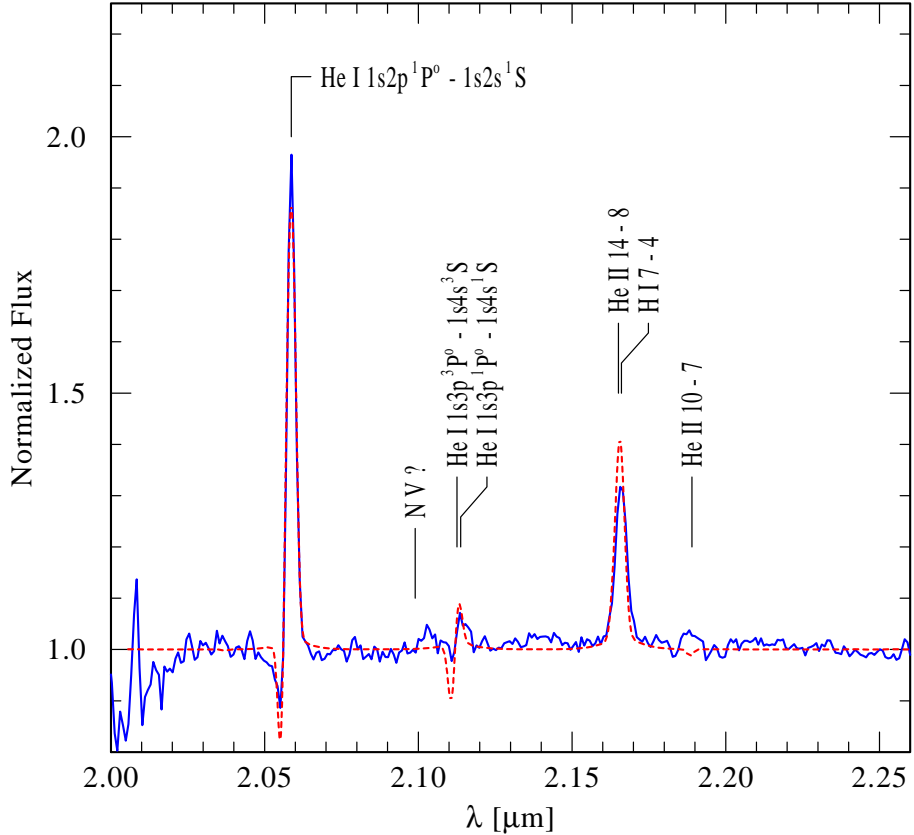}
\caption{Observed K-band spectrum of the Galactic Center star WR102ka (blue) and best fit model (red). The main lines are indicated. From \cite{barniske08}. } \label{barniske08_102}
\end{figure}

\underline{Wolf-Rayet stars}: J. Bestenlehner presented the analysis of Of/WN transition objects from a combined UV/optical/near-IR approach. Object VFTS 682 is very luminous but appears to be in isolation, questioning its formation process. Preliminary results on the analysis of O, Of/WN and WN stars indicate that Of/WN stars have wind properties in between those of O and WN stars. They also show a mass loss dependency on the Eddington factor, which indicate a larger L/M ratio for WN stars (Bestenlehner et al.\ in prep.)
W.R. Hamann presented studies of several WN and WC stars from multiwavelength and pure IR analysis. Of special interest is the so-called ``Peony'' star in the Galactic Center since its high luminosity makes it possibly the most massive Galactic star (\cite{barniske08}). Fig.\ \ref{barniske08_102} shows the fit of the K-band spectrum of that star. A WN star in the Scutum-Crux arm was also presented, highlighting the need for the J and H bands to correctly constrain the effective temperature (Burgemeister et al., in prep.).
P. Crowther focused on mid-infrared data of WC and WO stars obtained with \textit{Spitzer} and \textit{Herschel}. He highlighted the interest of forbidden lines to constrain stellar evolution. Such lines  being formed very far away from the photosphere (at heights from 10000 to 500000 stellar radius), they trace the outter wind. Ne abundances are derived in WC stars from \textit{Spitzer} spectra. A mass fraction of about 1\% is determined, in reasonable agreement with theoretical yields. In WO/WC stars, [OIII] 88.0 $\mu$m is used to constrain the oxygen content. From preliminary results on the binary $\gamma$ Vel (WC8+O), the O/C ratio seems to be a factor of two lower than the predictions of \cite{mm03}.

\smallskip

\underline{Red supergiants}: KM (red) supergiants: M. Bergemann presented analysis of red supergiants from pure near-infrared spectroscopy (\cite{bergemann12}). She raised the importance of non-LTE effects for the abundance determination of Fe, Ti, and Si and cautioned against using the TiO molecular fits for determination of effective temperature of RSG's, also because severe non-LTE effects in Ti ionization equilibrium will have an impact on the TiO formation. Corrections to LTE calculations have been quantified. Red supergiants are especially important for future telescopes/instruments since they are the brightest objects at infrared wavelengths.

\smallskip

\begin{figure}[]
\centering
\includegraphics[width=13cm]{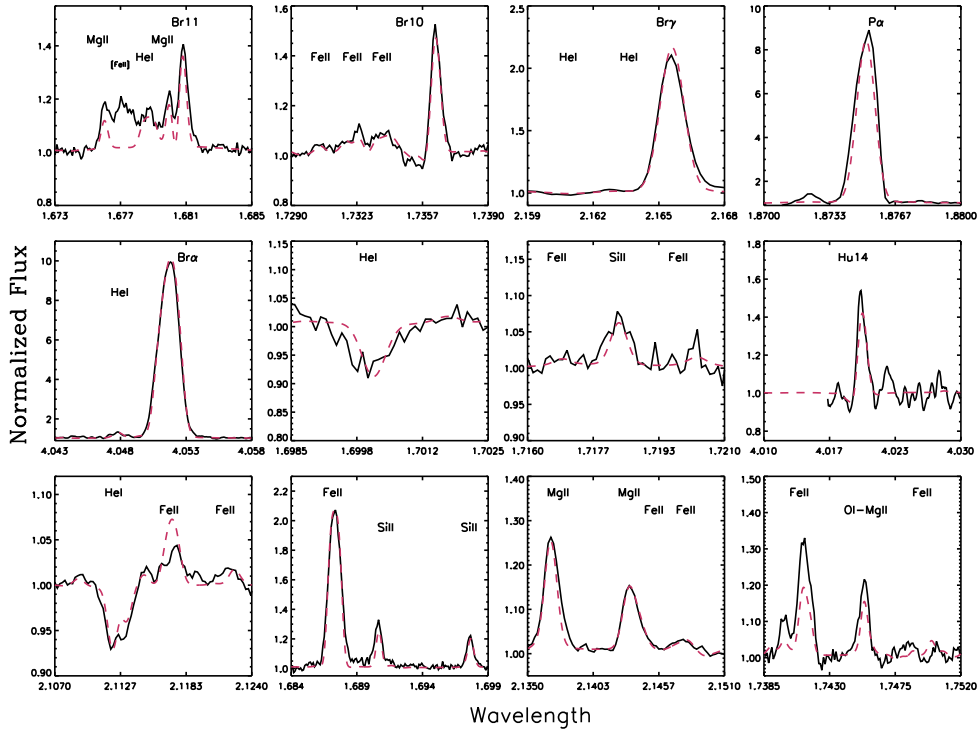}
\caption{Determination of Fe, Mg, Si, Na abundances in the LBV star Pistol. The black solid line is the observed spectrum, the red dashed line is th ebest fit model.  From \cite{paco09}. } \label{paco09_ab}
\end{figure}

\underline{Luminous Blue Variables}: N. Przybilla presented the results of \cite{paco09} on the surface chemical abundances of LBVs in the Quintuplet cluster. Najarro et al.\ were able to derive a solar Fe content, and an abundance in Mg, Si and Na about twice the solar abundance. LBVs are well suited for such analysis in the infrared since they display a number of metallic emission lines. This is illustrated in Fig.\ \ref{paco09_ab}. 

%-----------------------------------------------------------------------------------
\section{Conclusion}

Infrared phototemetry and spectroscopy can be used to determine stellar and wind properties of massive stars (hot and cool). The modelling of infrared spectra is more difficult than shorter wavelength spectra because the non-LTE effects are amplified. Accurate atomic data are necessary to correctly reproduce the observed line profile. Line-blanketing effects are also important. In spite of these difficulties, analysis based on pure infrared diagnostics usually give results consistent with optical studies. All types of massive stars can be studied. In addition to temperature, luminosity, gravity and mass loss rate, abundances of He, C, N, O, Mg, Si, Ne are feasible in cool stars and strong wind objects. Infrared studies are crucial to study stellar population in external galaxies with future extremely large telescopes.

\begin{acknowledgments}
FM thanks the french ``Agence Nationale de la Recherche'' for founding.

\end{acknowledgments}

\end{document}